# Overview of Cyber Science and Technology Programs at the U.S. Army Research Laboratory

Alexander Kott, Army Research Laboratory (ARL)

The U.S. Army Research Laboratory (ARL) received the first salvos in the battle for cybersecurity as early as three decades ago. In terms of technology history, it was an astonishingly long time ago. Before most people ever heard of the Internet. Before there were web browsers. Long before the smartphones. Back in 1986, the laboratory withstood attacks by Markus Hess, a Soviet-sponsored hacker who had successfully penetrated dozens of U.S. military computer sites. In his bestselling book, *The Cuckoo's Egg*, the pioneering U.S. cyber defender, Cliff Stoll, describes how he monitored the hacker's networks activities in the fall of 1986: "He then tried the Army's Ballistic Research Lab's computers in Aberdeen, Maryland. The Milnet took only a second to connect, but BRL's passwords defeated him: he couldn't get through" (Stoll 1989).

Two years later, the laboratory faced the legendary Morris Worm. "Around midnight on November 3, 1988, system managers at the Army's Ballistic Research Laboratory noticed their computers slowing down to a crawl as the worm stole precious computing processing time. Fearing a foreign attack, they pulled their computers off the nationwide network predating the Internet, called ARPAnet." (Hess, 2016)

The Army's Ballistic Research Lab, an ancestor of ARL, was the home of the ENIAC, the world's first electronic digital computer in 1946. It was also where the "ping" program was written in 1983, and where many other milestones of computing and networking took place. The encounters with the Soviet-sponsored hacker and with the Morris Worm were among such milestones.

Since those early beginnings, the history of ARL's efforts in cyber defense was exciting and challenging (Fig. 1). Although ARL is the Army's corporate laboratory that focuses on fundamental and early applied research (in the Department of Defense lingo – the research of 6.1 and early 6.2 types), the fundamental science endeavors are closely integrated with extensive operationally-oriented programs. These range from providing continuous cybersecurity defense services to multiple organizations, as well as cyber survivability and vulnerability analysis of Army systems.

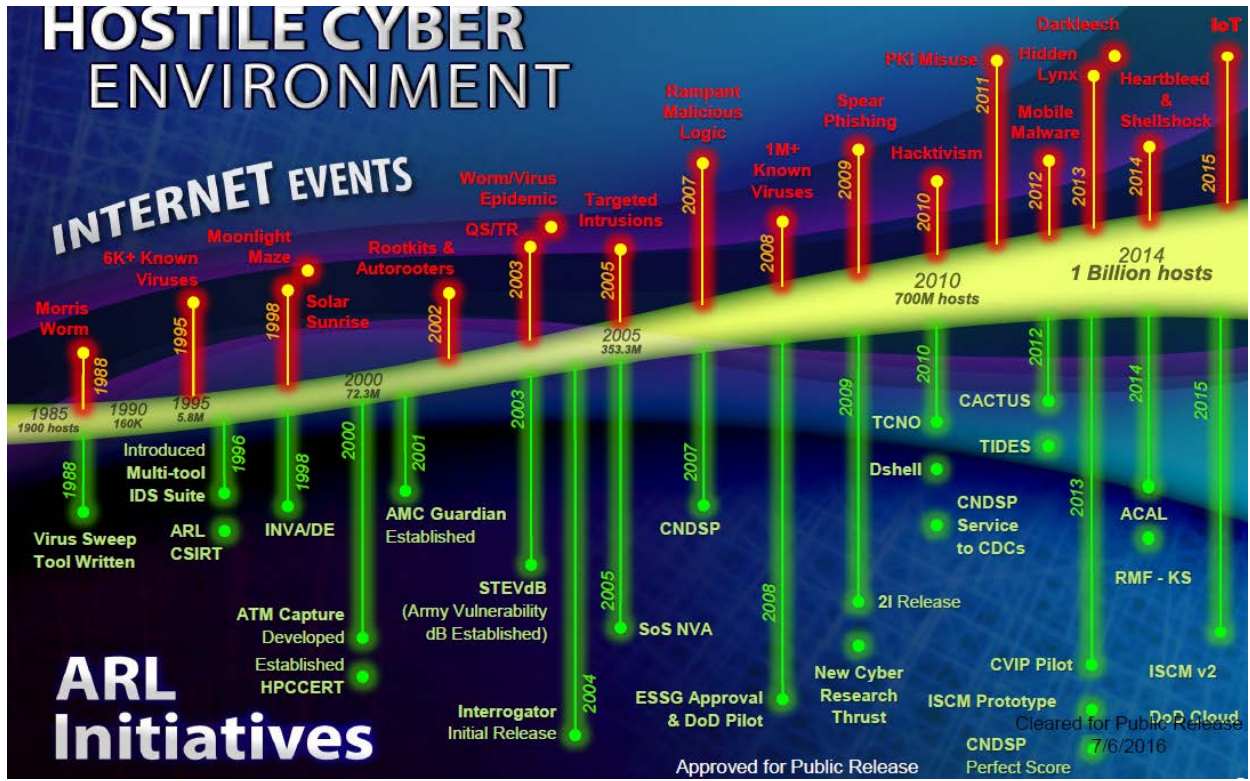

Fig. 1 ARL cyber research was informed by real-world environment

An important feature of ARL's business model is the great degree of collaboration with the academic community. One example is the Cyber Collaborative Research Alliance (CRA) (McDaniel and Swami 2016) that brings together, in closely integrated collaborative projects, ARL scientists with academic researchers from dozens of U.S. universities. Cyber CRA aims to develop the fundamental science of cyber detection, risk, agility, as well as the overarching challenge of human factors in cyber security. Similarly, the Network Science Collaborative Technology Alliance (see http://www.ns-cta.org/) integrates ARL and academic research efforts towards a broad understanding of how multi-genre networks of humans and information and communications devices influence each other and undergo complex dynamic transformations.

ARL collaborations are not limited to U.S. universities. ARL is also actively engaged with international partners.  ARL's Open Campus business model (http://www.arl.army.mil/www/default.cfm?page=2357) helps such wide-ranged collaborations by providing facilities and organizational support for enabling scientists and engineers from the U.S. and abroad to come to ARL for a period of time to work in partnership with ARL scientists.

Complementing its close ties with academic scientists, ARL research is also intertwined with practical, day-to-day operational responsibilities. Scientists are in direct communications with cyber analysts from the ARL Cybersecurity Service Provider (CSSP), a Tier II organization that defends networks of hundreds of customers belonging to all U.S. Military services, other government organizations, and even industrial entities (e.g.,  Oniha et al 2016). ARL has a strong reputation in the area of threat analysis and forensics.

The laboratory's experts in these fields are in high demand as they support cyber-related investigations conducted by law enforcement and counter-intelligence bodies. Vulnerability and survivability assessments of systems and networks that are either already deployed or are still in acquisition process, are another major area of ARL practical, hands-on contributions to Army cybersecurity (Acosta et al 2016). ARL's highly experienced teams of experts perform Cooperative Vulnerability and Penetration Assessments (blue team assessments) as well as Adversarial Assessments (red team assessments). Practical operational insights and needs obtained in operational activities are provided to scientists. They, in turn, utilize observations and data to develop new theories and models, and eventually to develop tools that transition into operational use.

Although participants of a broad cyber research community, ARL cyber scientists are largely driven by challenges unique to the ground operations of the Army. A key example is the exceptionally large attack surface of Army networks: the Army operates in environments within close proximity to allied and civilian assets and adversaries, comprising a complex cyber ecosystem. Forward-deployed network assets are vulnerable to cyber entry or physical capture and subversion of information and devices. Another distinct feature of Army cyber environments is the relatively disadvantaged assets, as the Soldiers' computing and communication devices are energy and weight constrained, with limited bandwidth and computational capacity. Large number of nodes and fast changes of Army cyber environments are also quite distinct. Soldiers operate in a mobile environment, in complex terrain, with rapidly changing connectivity. Lastly, these networks are often interspersed with civilian, allied, and adversarial networks.

These challenges inform and focus of ARL's research areas. One key area of research is the understanding the cyber threat. The topics in this area range from inferring influences and relations within a command and control organization from its encrypted communications, to novel uses of stylometry for identifying authors or origins of malware (Caliskan-Islam and Harang 2015), to tools and techniques for forensic analysis, and even to the study of cultural factors and personality that influence patterns of behaviors of cyber actors (Cho et al 2016).

Understanding the threats contributes to the characterization of risk experienced by a system or network. ARL's research in risk characterization includes such topics as statistical analysis of factors affecting the anticipated frequency of successful cyber attacks (Gil et al 2014), and theoretical approaches to network risk computation (Cam 2015). It also includes applied efforts to develop better procedures for risk inspection programs; tools for continuous monitoring of risk, cyber situational awareness (Kott et al 2014), and decision support systems for cyber risk assessments.

Knowing the risks helps focus the detection efforts (Kott and Arnold 2013). The comprehensive portfolio of ARL's research in detection of hostile cyber activities is based on close integration with practical network defense operations. It provides data and insights, and leads to the study of topics like the impact of packet loss in realistic cyber sensors on effectiveness of intrusion detection (Smith et al 2016). Other topics include special challenges of detection in cyber physical systems (Colbert 2016, Colbert and Kott 2016); use of machine learning for detection methods suitable for mobile, resource-constrained devices (Harang et al 2015, Harang 2016); cognitive models of human analyst's process of detection

(Acosta et al 2016); and synergistic approaches to human-machine intrusion detection (Ben-Asher and Yu 2016).

Ultimately, whether detected or not, the hostile cyber activities must be defeated. ARL explores approaches such as active cyber defense (Marvel et al 2014), post-intrusion triage for optimized recovery (Mell and Harang 2014), and cyber maneuvers that limit lateral propagation of hostile malware (Ben-Asher et al 2016).

These research projects are supported by a network of experimental facilities and laboratories dedicated to cyber research. For example, the ARL Cybersecurity Service Provider performs double duty: it supports large-scale operational cyber defense, but also acts as a laboratory for collection of real-world data for research, and a platform for insertion and testing of novel cyber defense tools continually invented and developed by ARL scientists.

Another example of a laboratory is the virtual laboratory called CyberVAN. It is an environment for design and execution of cyber experiments using virtual machines, real Army applications, and a network simulator capable of realistic portrayal of sizeable Army units in mobile operations in complex terrain.   CyberVAN is particularly well suited for experimental validation of theoretical results by academic researchers, including international collaborators.

Additionally, the Army Cyber-research and Analytics Laboratory at ARL serves as an environment that supports various industrial and federally-funded partners of ARL. Its functions range: personnel training, product integration, systems engineering, and integrated testing using real-world data. The CHIMERA laboratory specializes in the study of human factors and human-information interactions in cyber defense; it helps to explore the human dimension of cybersecurity.

All this research yields results, many of which transitioned to practice as tools and systems. For example, Interrogator is an ARL-developed suite of network monitoring, intrusion detection and intrusion analysis tools. Used at ARL, as well as at a number of other organizations, its architecture is optimized for government cyber security operations, for defense against sophisticated threats, and for rapid insertion of research tools as plug-ins. Another example, Interrogator-in-a-Box, was developed for defense of mobile tactical networks.  In addition, DShell is a framework for forensic analysis, popular with users at government agencies. ARL researchers attracted multiple, valuable international collaborators – and a good number of comments on social media – when they developed an open-source version of DShell and placed it on GitHub (see GitHub.com/USArmyResearchLab). Other examples of practical tools developed at ARL include COBWebS, a simulation tool that incorporate cyber warfare elements into training exercises, and a decision support tool for cybersecurity assessments, which helps perform assessments using public knowledge sources and custom data.

Looking further out, our long-term campaign of cyber research is guided by the vision of the future Army battlefield. In the year 2040, it will be a highly converged virtual-physical space, where cyber operations will be an integral part of the fight (Kott et al 2015). Cyber fires are the activities that will degrade, disrupt, deny, deceive and destroy not only informational, computational and communication resources of the adversary, but also the physical capabilities of its platforms, weapons, robots, munitions, and even of personnel. Cyber maneuver refer to activities that will rapidly move and transform the friendly informational-computational resources to deny the adversary an opportunity to attack, while imposing on him a new unsolvable problem (Fig. 2). Cyber fires and maneuver will rely on effective cyber intelligence collection capabilities.

Operating on multiple time scales, often far faster than human cognitive processes, in a highly dynamic, non-contiguous battlefield, these fires and maneuvers will join the conventional, kinetic fires and movements. Future cyber capabilities will have to support continuous (real-time, not just deliberate) planning and execution of highly agile, daring, aggressive cyber fires and maneuvers This will be performed in a way that is necessarily highly automated and reliant on machine intelligence, and yet responsive to human intent and guidance.

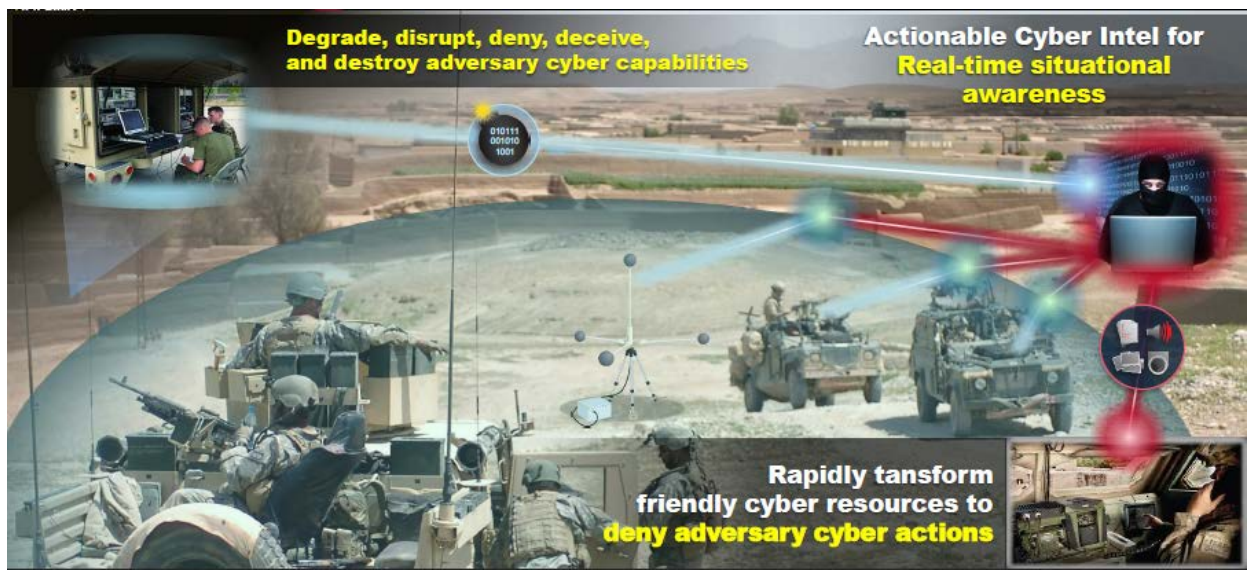

Fig. 2 ARL cyber research is increasingly focused on cyber fires and maneuvers in tactical environments

For these reasons, our cyber research efforts will increasingly focus on developing the models, methods, and understanding to overcome existing barriers to the realization of effective cyber fires and maneuvers in a tactical environment. The goals of this work are to pursue near-autonomous detection and identification of malicious activity directed at friendly networks; methods to rapidly respond to adversarial activities; predictive characterization of network vulnerabilities; and a robust framework to assess networks. Moreover, our research program will focus on the realization of methodologies for the reliable reconfiguration of friendly cyber assets to evade or recover from attack; covert means for collection and predictive analysis of enemy actions; and methodologies to degrade or destroy adversarial cyber assets with high certainty and predictable probabilities of success.

ACKNOWLEDGEMENTS: Iris Saunders helped prepare the manuscript, Jerry Clarke built the presentation that served as the outline of this article, and Latasha Solomon orchestrated the development of all articles.

REFERENCES

Acosta J, Edwards J, Shearer G, Parker T, Braun T, Marvel L. Modeling the decision processes of cybersecurity analysts to improve security assessments and defense strategies. Paper presented at: 23rd Annual National Fire Control Symposium (NFCS); 2016 Feb 8–11; Lake Buena Vista, FL.

Acosta, J.C., Padilla, E., Homer, J. and X. Ou. 2016. Risk analysis with Execution-Based Model Generation. Journal of Cyber Security and Information Systems 5(1):30-39

Ben-Asher N, Morris-King J, Thompson B, Glodek W. Attacker Skill, Defender Strategies, and the Effectiveness of Migration-Based Moving Target Defense in Cyber Systems. Paper presented at: 11th International Conference on Cyber Warfare and Security; 2016; Boston, MA.

Ben-Asher, N. and P. Yu. 2016. Synergistic Architecture for Human-Machine Intrusion Detection. Journal of Cyber Security and Information Systems 5(1):24-29

Caliskan-Islam A, Harang R, et al. De-anonymizing Programmers via Code Stylometry. SEC'15 Proceedings of the 24th USENIX Security Symposium; 2015; Washington, DC. Berkeley, CA: USENIX Association; c2015. p. 255-270.

Cam H. Risk Assessment by Dynamic Representation of Vulnerability, Exploitation, and Impact. In: Ternovakiy, IV, Chin P. Proc. SPIE 9458 Cyber Sensing; 2015 April 20-24; Baltimore, MD. SPIE Proceedings Vol. 9548; c2015.

Cho J, Cam H, Oltramari A. Effect of personality traits on trust and risk to phishing vulnerability: Modeling and analysis. 2016 IEEE International Multi-Disciplinary Conference on Cognitive Methods in Situation Awareness and Decision Support (CogSIMA). IEEE, 2016.

Colbert, E. 2016. Security of Cyber-Physical Systems. Journal of Cyber Security and Information Systems 5(1):40-47

Colbert, E.J. and Kott, A., 2016. Cyber-security of SCADA and other industrial control systems. *Advances in information security, 66*.

Ganin, A.A., Massaro, E., Gutfraind, A., Steen, N., Keisler, J.M., Kott, A., Mangoubi, R. and Linkov, I., 2016. Operational resilience: concepts, design and analysis. *Scientific reports*, *6*.

Gil, S., Kott, A. and Barabási, A.L., 2014. A genetic epidemiology approach to cyber-security. *Scientific reports*, *4*.

Harang R, Marvel L, Parker T, Glodek W. Bandwidth Conserving DCO Signature Deployment with Signature Set Privacy. IEEE MILCOM 2015; Tampa, FL; October 2015.

Harang, R. 2016. Machine Learning and Network Intrusion Detection: Results from Grammatical Inference. Journal of Cyber Security and Information Systems 5(1):18-23

Hess M. The Worm that Changed the Internet. Everything CBTN. From https://blog.cbtnuggets.com/2016/02/the-worm-that-changed-the-internet/. [accessed 2016 Feb 3].

Kott A, Alberts D A, Wang C. Will Cybersecurity Dictate the Outcome of Future Wars?. Computer 48.12 (2015):98-101.

Kott A, Arnold C. The promises and challenges of continuous monitoring and risk scoring. IEEE Security & Privacy 11.1 (2013):90-93.

Kott A, Wang C, Erbacher RF eds. 2014. Cyber Defense and Situational Awareness. New York: Springer.

Marvel L, Harang RE, Glodek WJ , Parker TW, Ritchey RP. A Proposed Model for Active Computer Network Defense. IEEE MILCOM 2014; Baltimore, MD; 2014 October.

McDaniel, P. and A. Swami. 2016. The Cyber Security Collaborative Research Alliance: Unifying Detection, Agility, and Risk in Mission-Oriented Cyber Decision Making. Journal of Cyber Security and Information Systems 5(1):10-17

Mell, P., & Harang, R. E. (2014, June). Using network tainting to bound the scope of network ingress attacks. In proceedings of the *Eighth International Conference on Software Security and Reliability* (pp. 206-215). IEEE.

Oniha, A., Weaver, G., Arnold, C. and T. Schreck. 2016. Information Security Continuous Monitoring (ISCM). Journal of Cyber Security and Information Systems 5(1):48-55

Smith SC, Hammell RJ, Parker TW, and Marvel LM. A theoretical exploration of the impact of packet loss on network intrusion detection. International Journal of Networked and Distributed Computing, 4(1): 2016 Jan 1.

Stoll C. The Cuckoo's Egg. New York, NY Simon & Schuster, 1989.